\documentclass[aps,prl,preprint,superscriptaddress,nofootinbib]{revtex4-2} 

\usepackage{chemformula} 
\usepackage{multirow} 
\usepackage{amssymb} 
\usepackage{enumitem} 
\usepackage[export]{adjustbox}
\usepackage{appendix}
\usepackage[font=small,labelfont=bf,tableposition=top]{caption}
\usepackage[utf8]{inputenc}
\DeclareUnicodeCharacter{0301}{\'}
\DeclareUnicodeCharacter{030C}{\v} 
\DeclareUnicodeCharacter{0308}{\"}
\DeclareUnicodeCharacter{0327}{\c}

\begin{document}

\title{Computational guide to optimize electric conductance in MoS\textsubscript{2} films}

\author{Alireza Ghasemifard}
    \email{alireza.ghasemifard@tu-dresden.de}
    \affiliation{TU Dresden, Theoretical Chemistry, Bergstr. 66c, 01062 Dresden, Germany}
    \affiliation{Helmholtz-Zentrum Dresden-Rossendorf, HZDR, Bautzner Landstr. 400, 01328 Dresden, Germany}
    \affiliation{Center for Advanced Systems Understanding, CASUS, Conrad-Schiedt-Straße 20, 02826 Görlitz, Germany}
\author{Agnieszka Kuc}
\email{a.kuc@hzdr.de}
    \affiliation{Helmholtz-Zentrum Dresden-Rossendorf, HZDR, Bautzner Landstr. 400, 01328 Dresden, Germany}
    \affiliation{Center for Advanced Systems Understanding, CASUS, Conrad-Schiedt-Straße 20, 02826 Görlitz, Germany}
\author{Thomas Heine}
    \email{thomas.heine@tu-dresden.de}
    \affiliation{TU Dresden, Theoretical Chemistry, Bergstr. 66c, 01062 Dresden, Germany}
    \affiliation{Helmholtz-Zentrum Dresden-Rossendorf, HZDR, Bautzner Landstr. 400, 01328 Dresden, Germany}
    \affiliation{Center for Advanced Systems Understanding, CASUS, Conrad-Schiedt-Straße 20, 02826 Görlitz, Germany}
    \affiliation{Yonsei University and ibs-cnm, Seodaemun-gu, Seoul 120-749, Republic of Korea}

\date{\today}

\begin{abstract}
Molybdenum disulfide (MoS\textsubscript{2}) is a high-potential material for nanoelectronic applications, especially when thinned to a few layers. 
Liquid phase exfoliation enables large-scale fabrication of thin films comprising single- and few-layer flakes of MoS\textsubscript{2} or other transition-metal dichalcogenides (TMDCs), exhibiting variations in flake size, geometry, edge terminations, and overlapping areas.
Electronic conductivity of such films is thus determined by two contributions: the intraflake conductivity, reflecting the value of each single layer, and charge transport across these overlapping flakes. 
Employing first-principles simulations, we investigate the influence of various edge terminations and of the overlap between flakes on the charge transport in MoS\textsubscript{2} film models. 
We identify characteristic electronic edge states originating from the edge atoms and their chemical environment, which resemble donor and acceptor states of doped semiconductors.
This makes either electrons or holes to majority carriers and enables selective control over the dominant charge carrier type (n-type or p-type).
Compared to pristine nanosheets, overlapping flakes exhibit lower overall conductance. 
In the best performing hexagonal flakes occurring in Mo-rich environments, the conductance is reduced by 20\% compared to the pristine layer, while the drop by 40\%, and 50\% is predicted for truncated triangular, and triangular flakes, respectively in S-rich environments.
An overlap of 6.5 nm is sufficient to achieve the highest possible interflake conductance. 
These findings allow for a rational optimization of experimental conditions for the preparation of MoS\textsubscript{2} and other TMDC semiconducting thin films.

\end{abstract}

\maketitle

\section*{Introduction}

Recent advancements in nanoelectronics have demonstrated the potential of semiconducting nanosheets, produced through liquid phase exfoliation (LPE), for the fabrication of high-performance printed transistors.\cite{kelly2017all, liu2023solution, kelly2022electrical, ippolito2023unveiling, ogilvie2022size} 
In this solvent-based technique, the single- to few-layered nanosheets inevitably exhibit overlap with varying alignments, stacking orders, and diverse edge termination.\cite{gabbett2024understanding}
While LPE is a powerful method to produce nanosheets, the resulting network exhibits a distribution in flake sizes, with mean lengths ranging from a few tens to hundreds of nanometers. \cite{backes2014edge}
Despite the well-established understanding of the intralayer electronic transport in two-dimensional (2D) semiconducting transition-metal dichalcogenides (TMDCs),\cite{gao2020tuning, lopez2013ultrasensitive, radisavljevic2013mobility, heine2015transition, kuc2011influence} a comprehensive investigation of electron transport across flakes at the atomistic scale remains necessary.
The interlayer transport at atomic scale provides the missing information needed to bridge the gap to macroscopic models and for the rational design of films with improved application-targeted transport properties.

The bottom-up approach of colloidal chemistry is a powerful method for synthesizing MoS\textsubscript{2} nanosheets with well-defined sizes. \cite{lin2015colloidal, oztas2014synthesis, son2016colloidal, zechel2023green}
Liquid cascade centrifugation (LCC) effectively separates these nanosheets into distinct size fractions. \cite{clifford2024emergent, backes2016production}
Nano-tomography provides means to acquire 3D images of nanosheets.\cite{gabbett2024quantitative}
Through engineering the edge composition of nanosheets, it becomes feasible to obtain MoS\textsubscript{2} flakes with a precise arrangement of Mo and S atoms. \cite{chen2017atomically, byskov1999dft, bollinger2003atomic}
It is possible to preferentially create zigzag (ZZ) edge configurations.\cite{wen2022mapping, lauritsen2007size, wen2022mapping, yang2018batch, tinoco2019metallic, cao2015role}
Importantly, the control over formation of edges primarily composed of ZZ-Mo or ZZ-S terminations,\cite{tinoco2019metallic, chen2017atomically, wang2014shape} opens up possibilities for tailoring the electronic properties of MoS\textsubscript{2} flakes.\cite{xu2017controllable, hakala2017hydrogen, kronberg2017hydrogen}
Raju et al. \cite{raju2020influence} recently showed that under Mo-rich conditions primarily hexagonal flakes with both ZZ-Mo and ZZ-S edges are formed.
Under S-rich conditions, the flake's morphology changes, gradually decreasing the ZZ-S edges and progressively expanding the ZZ-Mo-S\textsubscript{2} edges, resulting in a truncated triangular shape.
The transition to triangular flakes is driven by the increased stability of ZZ-Mo-S\textsubscript{2} edges at higher sulfur concentrations, which results in the complete suppression of ZZ-S edges.
MoS\textsubscript{2} with ZZ-Mo edges have been successfully synthesized via in situ heating techniques within a transmission electron microscope.\cite{wen2022mapping}
The formation of S-S dimers on Mo edges is promoted by highly sulfiding conditions at elevated temperatures. \cite{lauritsen2004atomic}

In this study, we investigate the influence of edge termination on the conductance of overlapping MoS\textsubscript{2} flakes, with a focus on how the overlapping region with different lengths, consisting of two stacked MoS\textsubscript{2} monolayers (MLs) with thermodynamically favorable H$_h^h$ stacking influences transport.
We focus on ZZ edges, which are the type of edge configurations in exfoliated MoS\textsubscript{2} flakes.
Armchair edges are energetically less favorable and difficult to obtain through exfoliation.
Using the non-equilibrium Green's function (NEGF) method,\cite{datta2016lessons} we calculate the electron transmission spectrum and quantum conductance.
Our findings show that: i) although overlapping flakes generally have lower conductance than pristine MLs, a greater overlap between layers, specifically a change from $\sim$~1 to 65~\AA~, increases quantum conductance from 1\% to 50\% and to 80\% (relative to the ML conductance) depending on the edge termination, ii) an overlap of 6.5 nm allows for maximum conductance between the flakes, iii) different edge states exhibit preferential charge carriers, favoring either donor- or acceptor-type behavior, leading to n- or p-type semiconductors, iv) ZZ-Mo edge (ZZ-S edge) demonstrate constructive (destructive) interference.
Our results should assist the rational design of more efficient electronic devices based on printed TMDC flake networks.

\section*{Models}

\begin{figure}[ht!]
\includegraphics[width=0.99\textwidth]{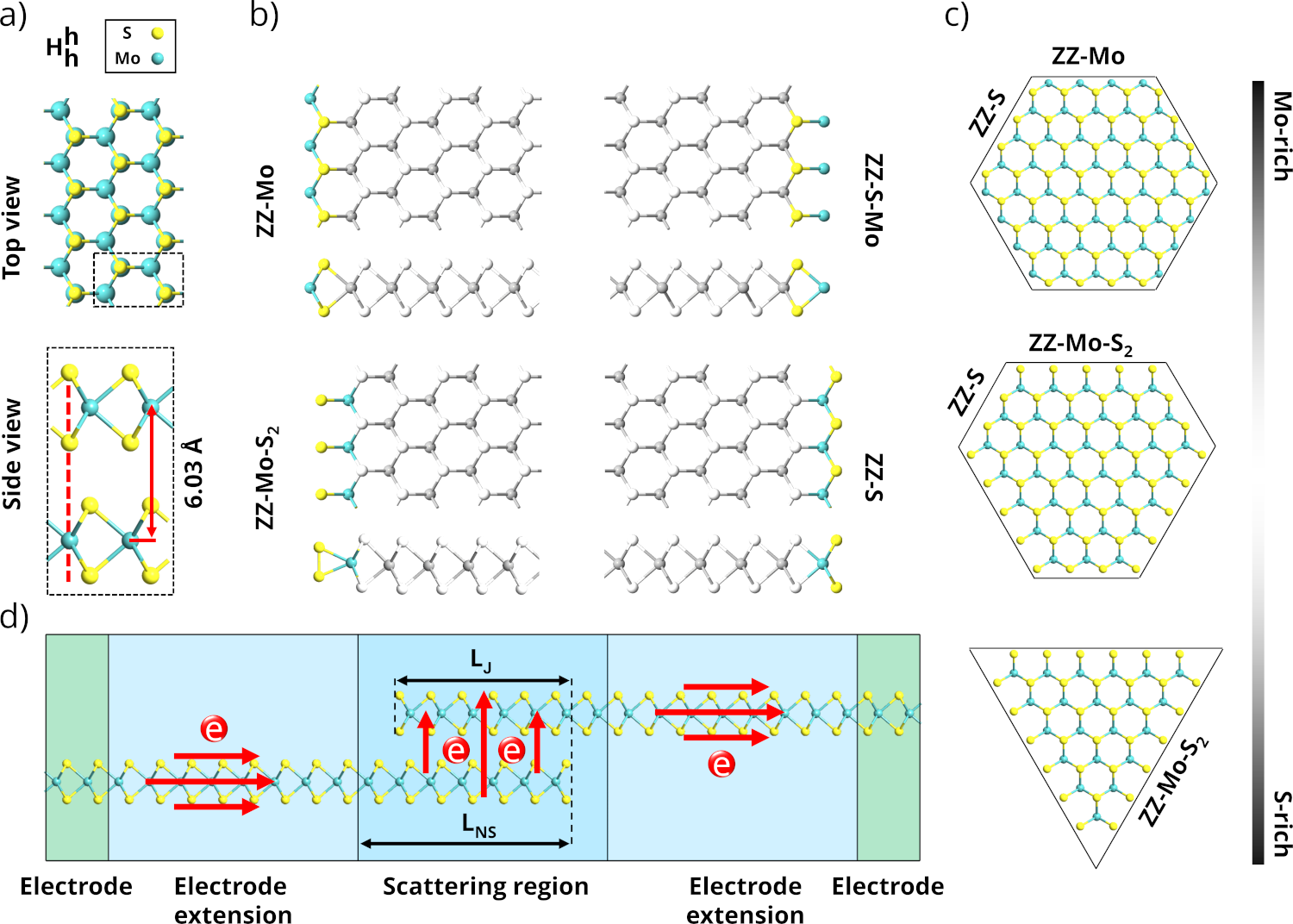}
\caption{Simulation models: (a) Top and side view of a bilayer (BL) MoS\textsubscript{2} in the $H_h^h$ polytype. Interlayer distance is indicated by the red arrow. Rectangular unit cell, necessary for electron transport simulations, is marked by the dashed lines. (b) Top and side view of possible ZZ-edges: pristine ZZ-Mo-edge, ZZ-S-Mo (S edge with Mo termination), ZZ-Mo-S\textsubscript{2} (Mo edge with S\textsubscript{2} dimer termination), and pristine ZZ-S-edge. (c) The schematic illustration of the relationship between flake edge and shape shows hexagonal with ZZ-Mo and ZZ-S edges. Truncated triangular with ZZ-S and ZZ-Mo-S\textsubscript{2} edges. Triangular with ZZ-Mo-S\textsubscript{2} edges. (d) Schematic representation of a device configuration. For the exemplary device, ZZ-S edge is used. The device consists of left and right electrodes and electrode extension, which are semi-infinite, and a scattering region. The two lateral and vertical transport paths are denoted by red arrows. $L_{NS}$ and $L_{J}$ are nanosheet and junction lengths, respectively, and the latter defines the overlap length.}
\label{fig_structure_device}
\end{figure}

We have investigated overlap MoS\textsubscript{2} MLs with the thermodynamically dominant H$_h^h$ stacking (Fig.~\ref{fig_structure_device} (a)) and different ZZ-edge terminations in individual layers (Fig.~\ref{fig_structure_device} (b)): pristine ZZ-Mo edge, ZZ-S-Mo (S edge with Mo termination), ZZ-Mo-S\textsubscript{2} (Mo edge with S\textsubscript{2} dimer termination), and pristine ZZ-S edge.
Fig.~\ref{fig_structure_device} (c) illustrates how the morphology of MoS\textsubscript{2} flakes evolves from hexagonal to triangular shapes with varying chemical species of Mo and S.
It also shows the relationship between flake edges and shapes: hexagonal flakes have three ZZ-Mo and ZZ-S edges, truncated triangular flakes have ZZ-S and ZZ-Mo-S\textsubscript{2} edges, and triangular flakes have only ZZ-Mo-S\textsubscript{2} edges. 
We simulated a total of four device models of overlap MoS\textsubscript{2} MLs with the above-mentioned edge terminations and with varying overlap lengths, L$_J$ (Fig.~\ref{fig_structure_device} (d)).
While devices with mixed edge terminations of the two layers are also possible, they are beyond the scope of the present work.

A model device consists of the left and right semi-infinite electrodes, composed of MoS\textsubscript{2} MLs, and the central transport region, modeled as overlapping MoS\textsubscript{2} MLs.
The transport channel (also called scattering region) is non-periodic along the transport direction (from source to drain) and consists of varying $L_{J}$, as shown in Supplementary Information (SI) Fig.~\ref{fig_SI1}.
The in-plane direction normal to the transport is periodic.
A vacuum of 20~\AA\ in the out-of-plane direction was used to avoid any interactions caused by periodic images.

Details of the computational procedures employed, including structural optimization, electronic structure calculations (surface band structure and device density of states (DDOS)), and quantum conductance simulations, are provided in the METHODS.
Note that the present approach allows simulations of transport within the coherent limit, neglecting inelastic scattering mechanisms, such as electron-phonon interactions.

\section*{Results and Discussion}

\begin{figure}[ht!]
    \centering
    \includegraphics[width=0.8\textwidth]{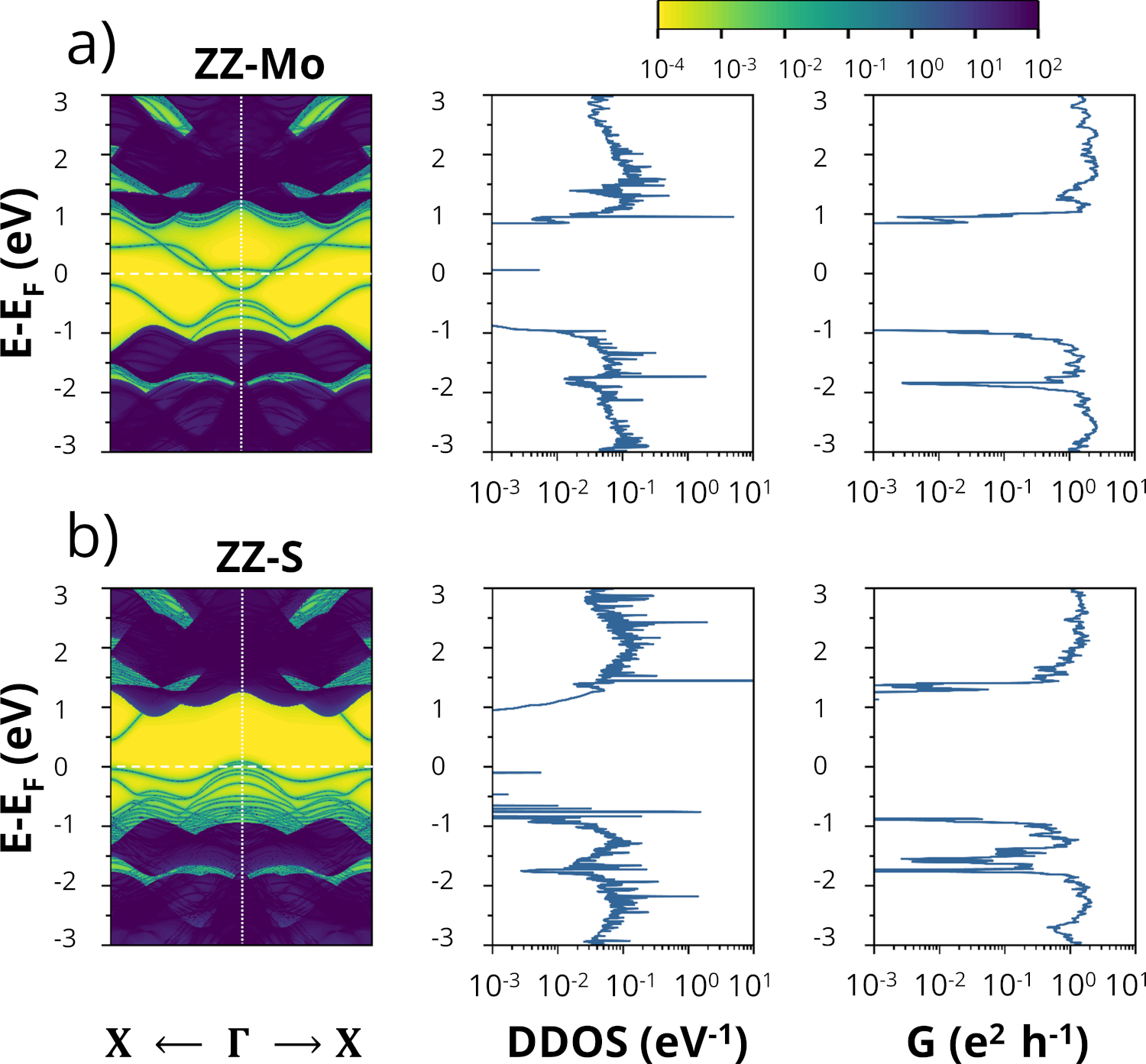}
    \caption{Surface band structure, device density of states, and conductance as function of energy for (a) ZZ-Mo and (b) ZZ-S model. The Fermi level is shifted to zero (white dashed horizontal line). The color legend denotes weighted device DOS in $eV^{-1}$. Dark blue - bulk states, light green with yellow background - edge states.}
    \label{fig_band_structure}
\end{figure}

The simulated electronic properties of ZZ-Mo and ZZ-S shown in Fig.~\ref{fig_band_structure} (see Figs.~\ref{fig_SI2}, and \ref{fig_SI3} for ZZ-S-Mo, and ZZ-Mo-S\textsubscript{2} respectively) are the surface band structure, DDOS, and conductance ($G$) as function of energy.
In addition to the bulk states (dark color), the surface band structure shows the presence of electronic edge states (light green color).
A free charge carrier's response depends on $\Delta E = |E_{e} - E_{VBM, CBM}|$, the energy difference between the relevant transport state, i.e., the conduction band minimum (CBM) for electrons or the valence band maximum (VBM) for holes, and the edge state ($E_{e}$).
States within $\Delta E < 4k_{B}T$ near the band edges, where $k_{B}$ is Boltzmann constant, and $T$ is the temperature, act as dopants, which can restrict the movement of free carriers. \cite{jin2020s, panarella2024evidence}. 

\begin{figure}[ht!]
    \centering
    \includegraphics[width=0.99\textwidth]{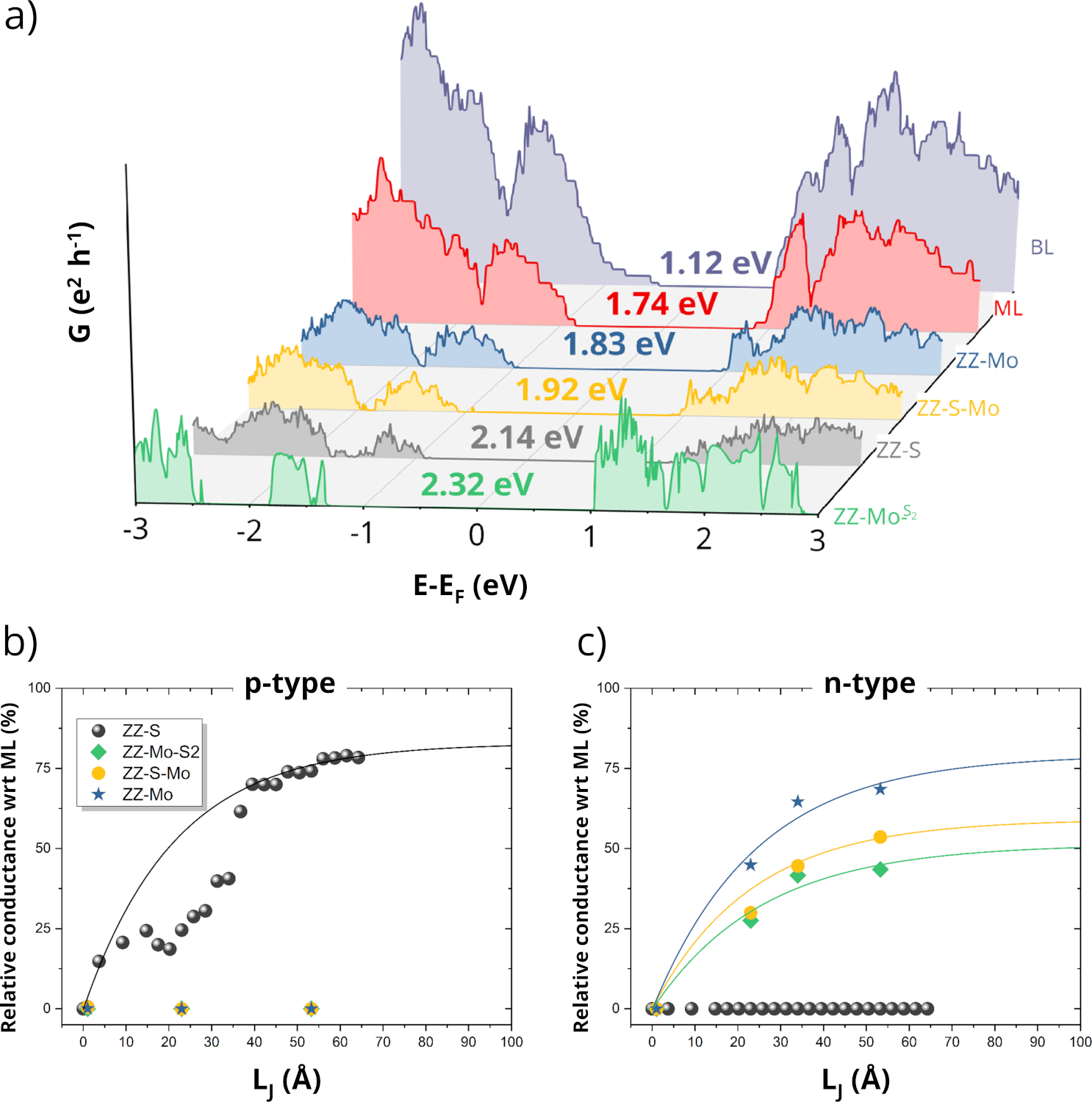}
    \caption{(a) Conductance as function of energy for all four studied devices, compared to ML and BL. The transport gap of each device is denoted by an arrow and given with color code. The Fermi level is shifted to zero. (b) Relative quantum conductance over Fermi distribution function with respect to ML for the states in a $\pm4k_{B}T$ energy window around VBM, and (c) in a $\pm4k_{B}T$ energy window around CBM of ML (see Fig.~\ref{fig_SI4}). Quantum conductance increases as the $L_J$ increases and reaches saturation at 50\%, 60\%, 80\%, and 80\% of the ML value in ZZ-Mo-S\textsubscript{2}, ZZ-S-Mo, ZZ-S, and ZZ-Mo, respectively, with $L_J$ $\geq$ 6.5 nm.}
    \label{fig_all_G_L}
\end{figure}

Devices with ZZ-Mo, ZZ-S-Mo, and ZZ-Mo-S\textsubscript{2} exhibit a high concentration of edge states near CBM, whereas ZZ-S exhibits them near VBM.
In ZZ-Mo, ZZ-Mo-S\textsubscript{2}, and ZZ-S-Mo the edge states are likely unoccupied, particularly around the $\Gamma$ point, and act predominantly as donors, resulting in n-type semiconductors, as evidenced in Figs.~\ref{fig_band_structure} (a), ~\ref{fig_SI2} (a), and \ref{fig_SI3} (a). 
In contrast, in ZZ-S, the edge states are more likely occupied and behave as acceptors, leading to a p-type semiconductor, as shown in Fig.~\ref{fig_band_structure} (b).
A summary of transport properties of all devices is shown in Fig.~\ref{fig_all_G_L} (a).
Among the investigated devices, the largest conductance (and smallest transport gap) was obtained for ZZ-Mo, while the lowest conductance for ZZ-S, and largest transport gap for ZZ-Mo-S\textsubscript{2} all with $L_{J}$ = 23~\AA\ (cf. Fig.~\ref{fig_all_G_L} (a)).

Apart from the edge states, a change in the overlap length $L_J$ of the stacked layers influences the transport.
Quantum conductance increases with the systematic increase of $L_J$ as depicted in Fig.~\ref{fig_all_G_L} (b) and (c).
$L_J$ is increased in steps of half the lattice vector (2.75~\AA) along the transport direction, see Fig.~\ref{fig_SI1}.
The increase in conductance can be attributed to the increased DOS at the interface, allowing more electrons to be transported. 
The conductance over a Fermi distribution function in a $\pm4k_{B}T$ energy window around CBM and VBM of ML (see Fig.~\ref{fig_SI4}), supports the p-type characteristic for ZZ-S and the n-type characteristic for the other edge configurations.
As $L_{J}$ continues to increase beyond 6.5 nm, the quantum conductance enters the saturation region and remains constant thereafter. 
Beyond 6.5 nm, the quantum conductance depends on the specific edges with both ZZ-S and ZZ-Mo approaching $\approx$ 80\% conductance of ML at large overlaps, as shown in Fig.~\ref{fig_all_G_L} (b) and (c). 
ZZ-Mo-S\textsubscript{2} and ZZ-S-Mo approaching 50\% and, $\approx$ 60\% conductance of ML at large overlaps, respectively (cf. Fig.~\ref{fig_all_G_L} (c)).
Therefore, while the ML semiconductor provides a reference conductance, the edge configuration and the nanosheet length $L_{NS}$ determine the overall transport properties of the stacked structures.
Generally, we notice that the size of the transport gap exhibits an inverse dependence on the $L_{J}$, gradually decreasing as the overlap length increases (see Fig.~\ref{fig_SI5}).

Also the interlayer distance ($d$) between overlapping layers influences device performance, with conductance increasing (decreasing) as $d$ decreases (increases).
A compression of 1~\AA\ from the equilibrium interlayer distance results in a conductance enhancement of 27\% and 22\% for ZZ-Mo and ZZ-S, respectively. 
In contrast, expansion by 1 \AA\ leads to a decrease in conductance by 40\% and 50\% in ZZ-Mo and ZZ-S, respectively (cf. Fig.~\ref{fig_SI6}).

\begin{figure}[ht!]
    \centering
    \includegraphics[width=0.98\columnwidth]{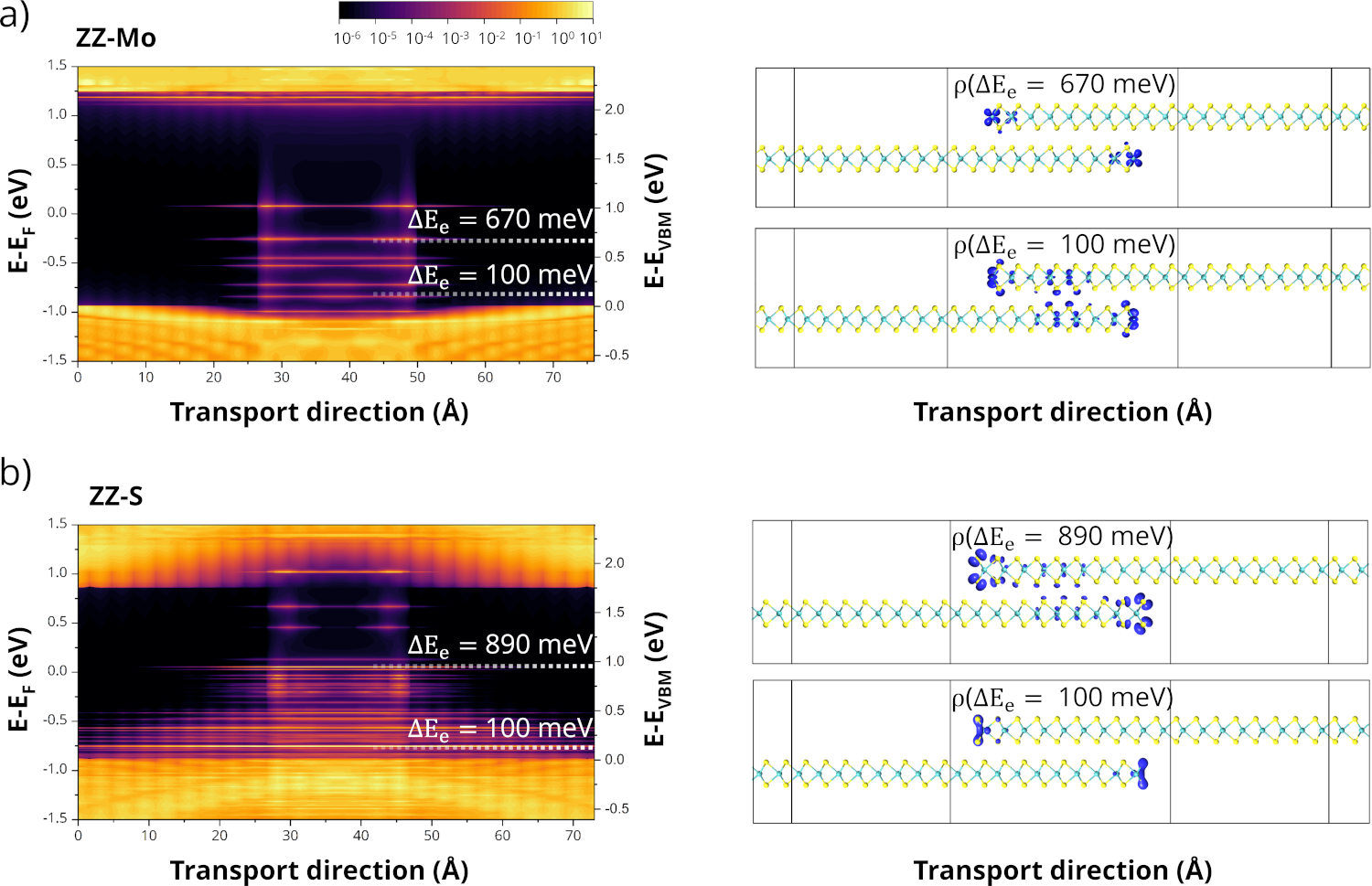}
    \caption{Local density of states as function of position (left panels) for (a) ZZ-Mo and (b) ZZ-S, along with the edge states' wave function spread in real space (right panels). The color scale represents the DDOS in units of eV$^{-1}$. The Fermi level is shifted to zero and shown on the left axis.}
    \label{fig_PLDOS}
\end{figure}

To investigate the effect of edges and overlap on conductance, we examined the electronic properties in more detail.
Figs.~\ref{fig_PLDOS} and \ref{fig_SI7} show the local DOS (LDOS) for each device, with the corresponding wave function in real space.
Depending on the type of the edge, different atomic orbitals dominate in the wavefunction: in ZZ-Mo, states with high $\Delta E$ are formed from Mo-d\textsubscript{xy} orbitals, while Mo-d\textsubscript{z2} orbitals are responsible for the states with smaller $\Delta E$.
In ZZ-S, S-p\textsubscript{z} and -s orbitals form the states with high and small $\Delta E$, respectively.
In the cases with mixed Mo-S or S-Mo edges, both S-p\textsubscript{z} and Mo-d\textsubscript{z2} orbitals contribute to the LDOS.

\begin{figure}[ht!]
    \centering
    \includegraphics[max size={\textwidth}{0.6\textheight}]{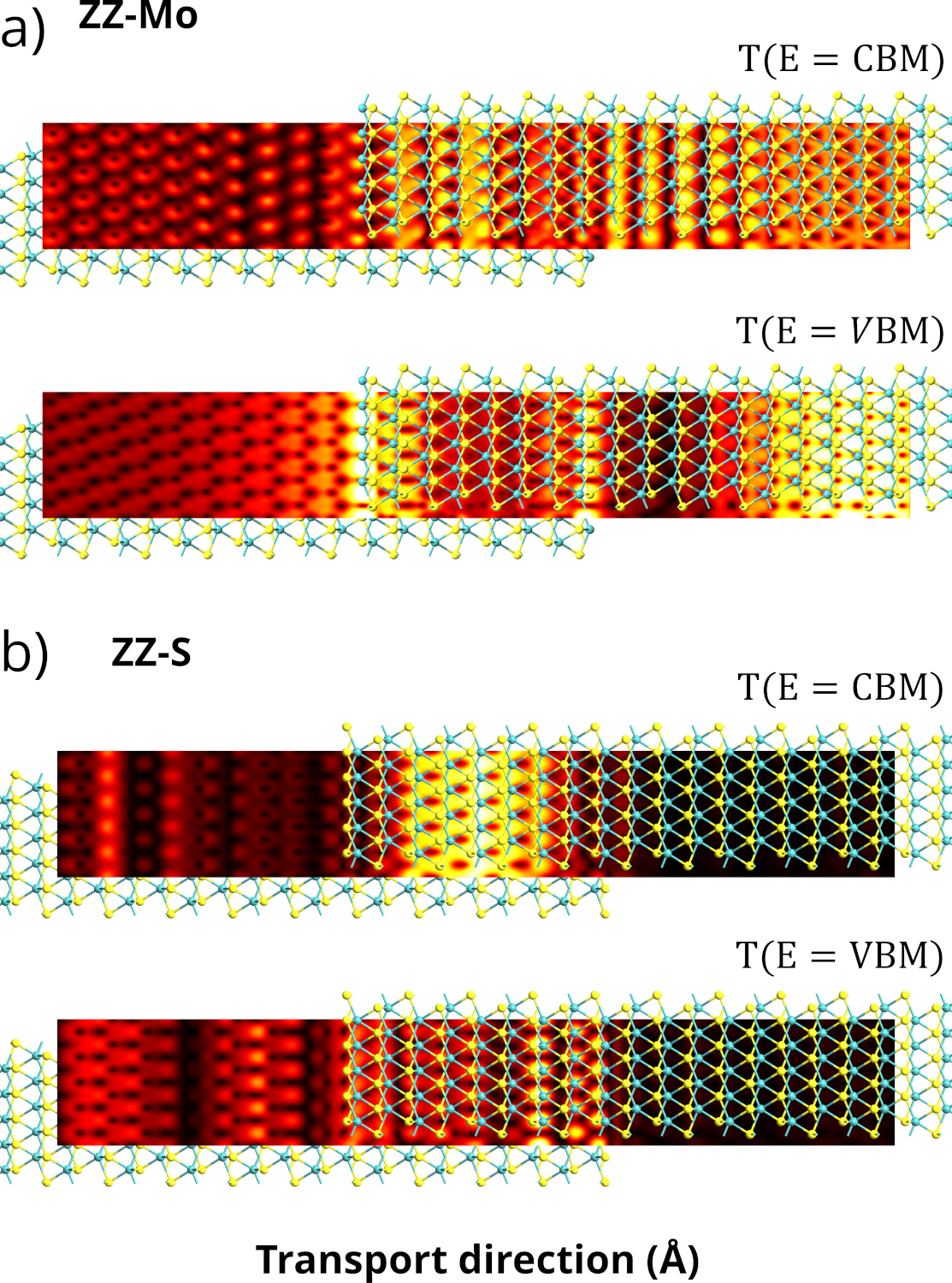}
    \caption{Transmission eigenstates of (a) ZZ-Mo with constructive interference and (b) ZZ-S with destructive interference both at CBM and VBM energies, and $k\textsubscript{a}=\Gamma$. The calculated transmission eigenstates have the highest eigenvalue. The amplitude of the transmission eigenstates is in the order of eV$^{-1/2}$~\AA$^{-3/2}$. The bright area has the maximum and the dark area has the minimum amplitude.}
    \label{fig_eigenstates}
\end{figure}

Analyzing transmission eigenstates at both VBM and CBM at the $\Gamma$ point, we observe constructive transmission states (bright regions) throughout the entire ZZ-Mo device, as shown in Fig.~\ref{fig_eigenstates} (a).
A change from ZZ-Mo to ZZ-Mo-S\textsubscript{2} under S-rich conditions preserves the constructive interference pattern (see Fig.~\ref{fig_SI3} (b)) with a decrease in conductance by 18\% with $L_{J}$=23~\AA\ and 30\% with $L_{J}$=65~\AA, indicating that the presence of S\textsubscript{2} termination in ZZ-Mo-S\textsubscript{2} has more impact on the overall transport properties in large overlap lengths.
In contrast, the right part of the ZZ-S device exhibits destructive transmission states (dark regions), implying that there is no effective electron transport between the layers (see Fig.~\ref{fig_eigenstates} (b)).
In ZZ-S-Mo, a different behavior was observed, such as constructive transmission at CBM states and destructive transmission at VBM states, as shown in Fig.~\ref{fig_SI2} (b), which supports the n-type characteristics of ZZ-S-Mo.

In a 3D network of 2D nanosheets composed of many well-defined, overlapping flakes, the total resistance can be expressed as  $R_{\mathrm{Net}} = R_{\mathrm{NS}} + R_{\mathrm{J}}$, where $R_{\mathrm{NS}} \propto L_{\mathrm{NS}}$ represents the resistance of an individual nanosheet (without overlap) and $R_{\mathrm{J}} \propto L_{\mathrm{J}}$ is the resistance of the junction formed by the overlap. 
As charge carriers traverse the network, they inevitably cross these overlapping junctions, encountering resistance that varies depending on the edge types of the constituent nanosheets. 
The overall trend shows that hexagonal flakes are the most interesting models for efficient interlayer transport. 
Therefore, in order to have maximum efficiency of a device, ZZ edges with either Mo or S are preferred.

\section*{Conclusion}

This study provides a computational guide to optimize electrical conductance in MoS\textsubscript{2} films made of overlapping flakes of the commonly observed flake shapes and edge terminations.
We show that besides the flake size, the degree of overlap between flakes, the edge termination, and the interlayer spacing are important factors for correctly assessing the transport properties.
Our first-principles calculations demonstrate that hexagonal flakes occurring in Mo-rich environments with overlapping ZZ-Mo or ZZ-S edges can achieve the best transport performance.
Truncated triangular and triangular flakes with overlapping ZZ-Mo-S\textsubscript{2} edges occurring in S-rich environments show poor transport performance.
Overlapping flakes decrease overall conductance compared to pristine nanosheets, with conductance drops ranging from 20\% in the best-performing system to 50\% in the poor-performing system.  
Increasing the flake concentration to achieve an overlap of more than 6.5 nm does not improve conductance. 
ZZ-Mo, ZZ-Mo-S\textsubscript{2}, and ZZ-S-Mo edges exhibit constructive interference at donor states, favoring n-type semiconductors.
Conversely, ZZ-S edge displays destructive interference at acceptor states with p-type semiconductors.
Compressing the interlayer distance at the flake overlap interfaces by 1~\AA\ from the equilibrium distance results in a conductance enhancement of 27\% for ZZ-Mo and 22\% for ZZ-S, while a similar expansion decreases conductance by 40\% and 50\% for ZZ-Mo and ZZ-S, respectively. 
Thus, a controlled process is needed to construct optimized MoS\textsubscript{2} films with controlled flake geometry and overlaps for achieving the best electrical properties.

\section*{Methods}
\subsection*{Details of computational methods}
All systems were fully relaxed using LAMMPS code\cite{LAMMPS} with Reax~\cite{ReaxFF1_vanDuin2001, ReaxFF2_chenoweth2008, ReaxFF3_aktulga2012, ostadhossein2017reaxff} force field (ReaxFF) with a maximum force component of 0.1\,eV/\AA\ and pressure of 10\,kbar, as obtained from our previous work on BL MoS\textsubscript{2}.\cite{arnold2023relaxation, arnold2024implementing}
The atomic positions were subsequently reoptimized with ReaxFF after creating different types of edges to allow for edge reconstruction.
To accurately model the formation of S dimers at the edge of ZZ-Mo-S\textsubscript{2}, a process not inherently captured by the ReaxFF parameterization (see Fig.~\ref{fig_SI8}), further optimization was carried out using density functional theory (DFT).  
DFT optimization was performed as implemented in QuantumATK S-2022.03 package,\cite{smidstrup2020quantumatk} using the Perdew–Burke–Ernzerhof (PBE) functional,\cite{perdew1996generalized} vdW D3 correction,\cite{grimme2011effect} and double-zeta polarized pseudopotentials.

Electronic structure calculations (surface band structure and device density of states - DDOS) were performed employing DFT method with PBE functional and pseudo-Dojo pseudopotentials.\cite{van2018pseudodojo}
DDOS represents the contribution of a device's left and right electrodes projected onto the atoms within its central region. 
Surface band structure is essentially DDOS mapped along a route of k-points. 
This allows us to differentiate edge states from bulk states.
A dense $11\times1\times199$ in a- (periodic), b- (vacuum), and c-direction (transport) Monkhorst-Pack k-point grid and $45$ hartree density mesh cutoff were used.
DOS was calculated using the tetrahedron method,\cite{blochl1994improved} in order to capture the fine feature of the electronic edge states. 
The quantum conductance was calculated for each device using DFT with PBE functional in combination with NEGF and Landauer-B\"uttiker (LB) approach:\cite{buttiker1985generalized, datta2016lessons, datta2018lessons}
\begin{equation}
    G(E)=\frac{2e^2}{h}tr\left[ \mathcal{G}^\dagger\Gamma_{R} \mathcal{G}\Gamma_{L} \right]
\end{equation}
\noindent where $e$ and $h$ are fundamental constants, $\mathcal{G}$ is the total Green's function (GF) and, 
\begin{equation}
    \Gamma_{\alpha}=i\left[ \Sigma_{\alpha}-(\Sigma_{\alpha})^\dagger \right]    
\end{equation}
\noindent are the coupling matrices, where $\Sigma_{\alpha}$ is the self-energy, with $\alpha=L, R$, corresponding to the left and right electrodes.

\section*{Acknowledgments}
This project has received funding from the European Union’s Horizon 2020 research and innovation program under grant agreement No 956813. The authors also thank Deutsche Forschungsgemeinschaft (DFG) for funding within CRC 1415 and SPP 2244 projects.
The authors gratefully acknowledge the computing time provided to them on the high-performance computers Barnard and Noctua 2 at the NHR Centers NHR@TUD (ZIH) and NHR@Padeborn (PC\textsuperscript{2}).

\section*{ORCiD IDs}
\noindent\href{https://orcid.org/0000-0003-0059-3814}{Alireza Ghasemifard https://orcid.org/0000-0003-0059-3814} \\
\href{https://orcid.org/0000-0002-9458-4136}{Agnieszka Kuc https://orcid.org/0000-0002-9458-4136} \\
\href{https://orcid.org/0000-0003-2379-6251}{Thomas Heine https://orcid.org/0000-0003-2379-6251}

\section*{TOC figure}
\begin{center}
    \includegraphics[width=1.0\linewidth]{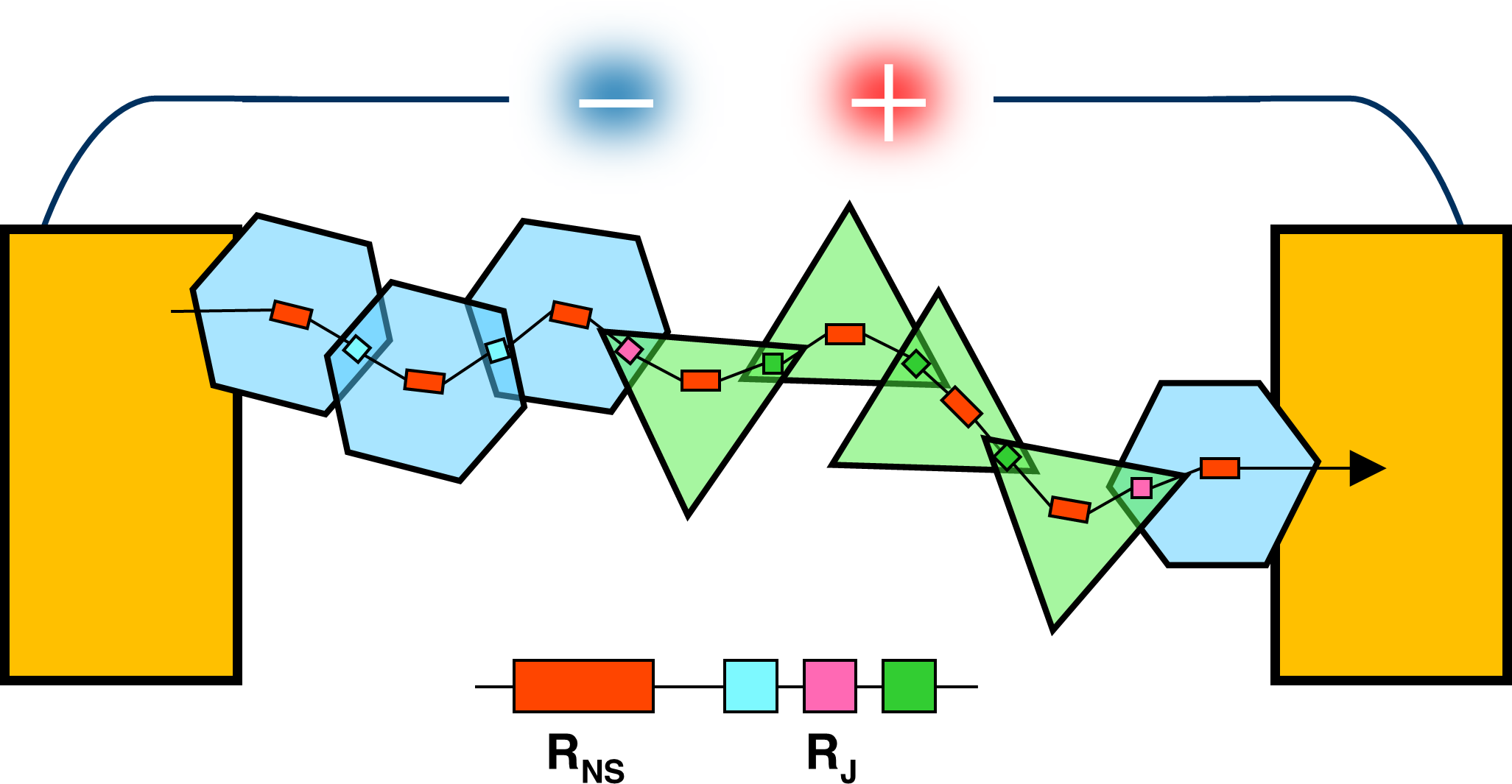}
\end{center}

\section*{REFERENCES}
\providecommand{\latin}[1]{#1}
\makeatletter
\providecommand{\doi}
  {\begingroup\let\do\@makeother\dospecials
  \catcode`\{=1 \catcode`\}=2 \doi@aux}
\providecommand{\doi@aux}[1]{\endgroup\texttt{#1}}
\makeatother
\providecommand*\mcitethebibliography{\thebibliography}
\csname @ifundefined\endcsname{endmcitethebibliography}
  {\let\endmcitethebibliography\endthebibliography}{}

\clearpage
\appendix
\section*{SUPPLEMENTARY INFORMATION}

\setcounter{figure}{0}
\renewcommand{\figurename}{FIG.}
\renewcommand{\thefigure}{S\arabic{figure}}

\begin{figure}[ht!]
    \centering
    \includegraphics[max size={\textwidth}{0.8\textheight}]{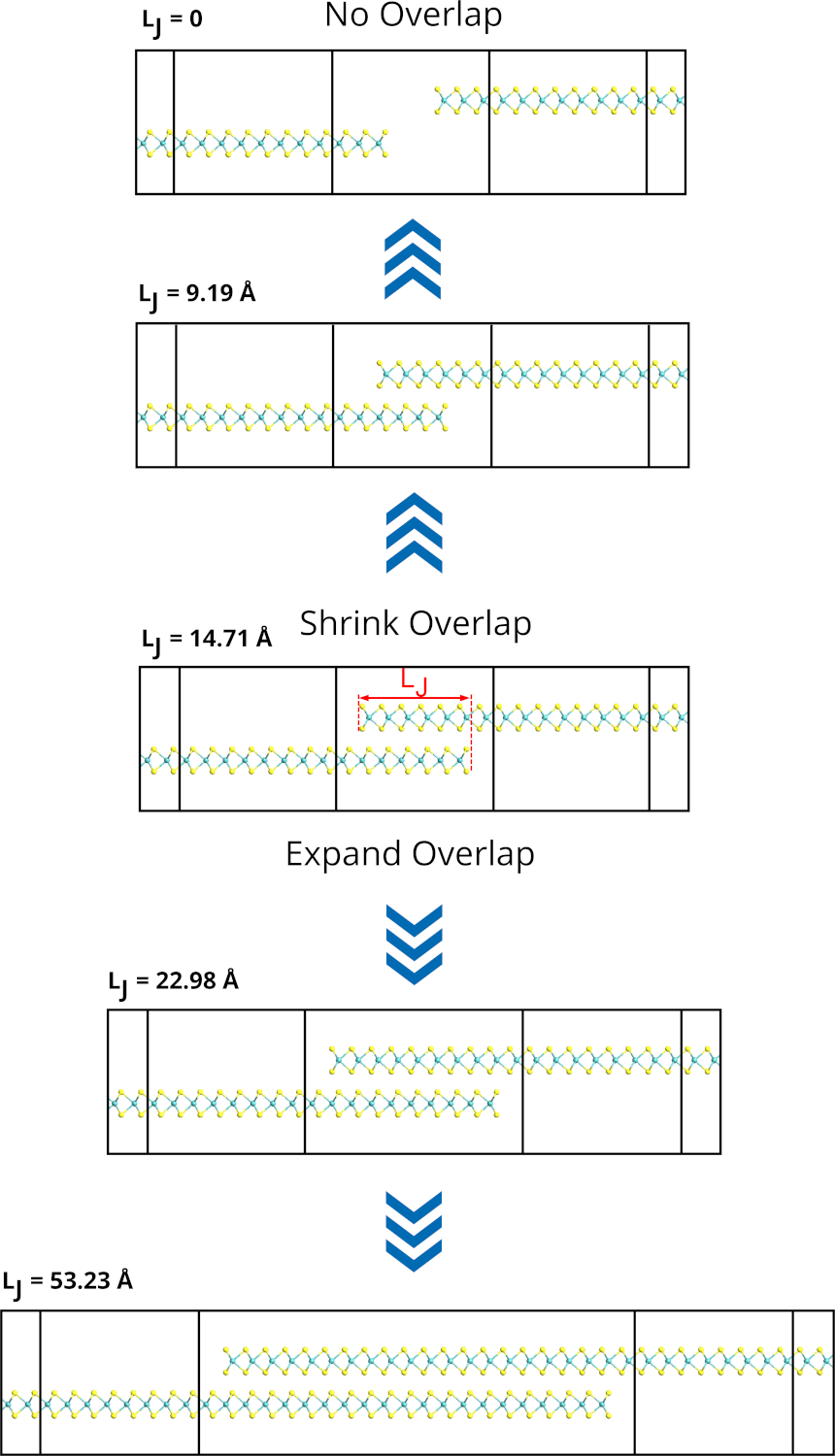}
    \caption{Schematics of change in overlap length of a device. Starting with the exemplary ZZ-S device with $L_{J}$=14.71~\AA, it is possible to adjust the overlap length of the scattering region by removing or adding a MoS\textsubscript{2} unit in each layer. This allows contraction or expansion of the scattering region, respectively. The left and right leads, as well as the extension leads, remain unchanged.}
    \label{fig_SI1}
\end{figure}

\begin{figure}[ht!]
    \centering
    \includegraphics[width=0.7\textwidth]{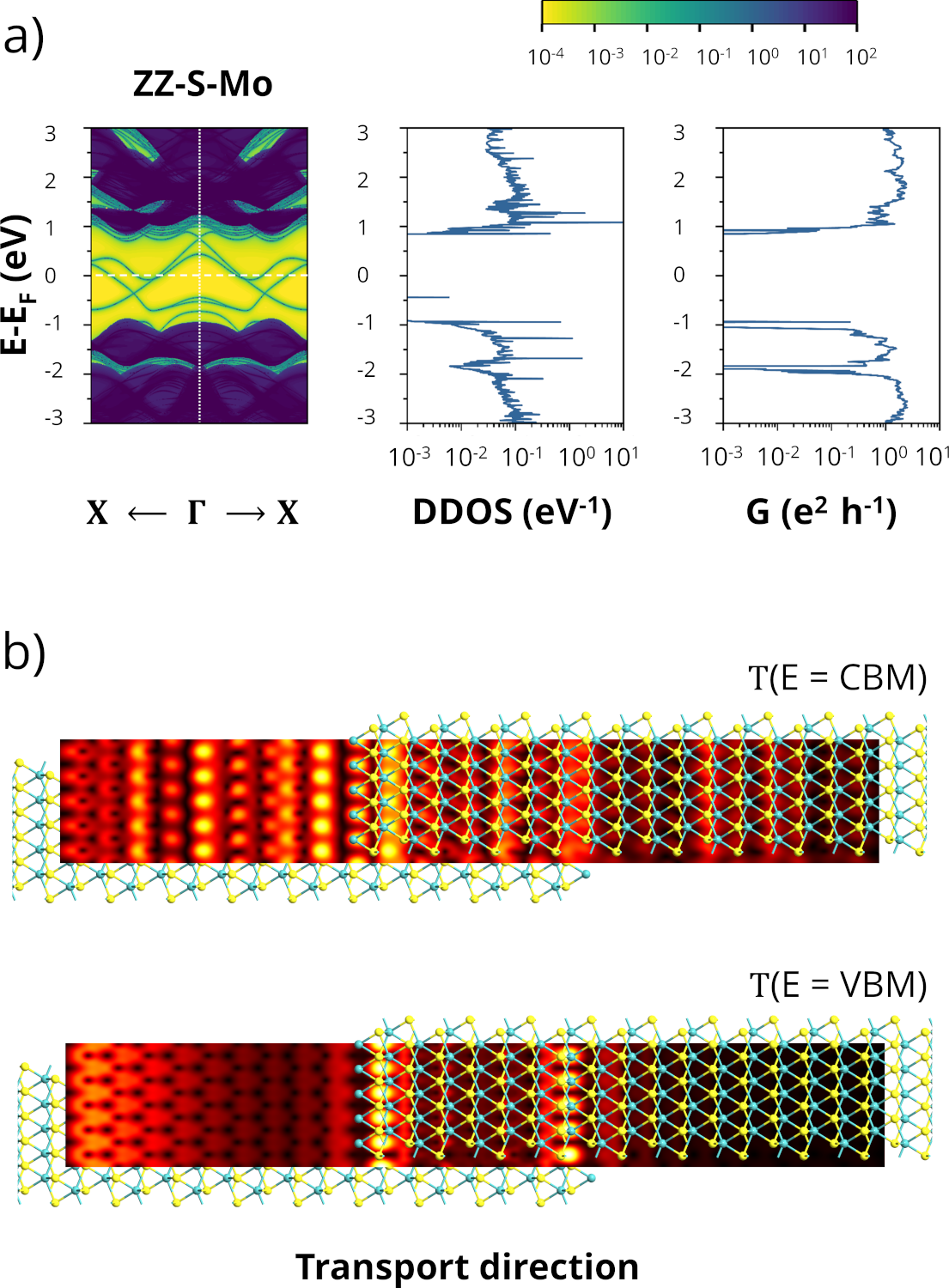}
    \caption{(a) Surface band structure, device density of states, and conductance as function of energy. (b) Transmission eigenstates at $E=CBM$ and $E=VBM$, with $k\textsubscript{a}=\Gamma$ for ZZ-S-Mo device.}
    \label{fig_SI2}
\end{figure}

\begin{figure}[ht!]
    \centering
    \includegraphics[width=0.7\textwidth]{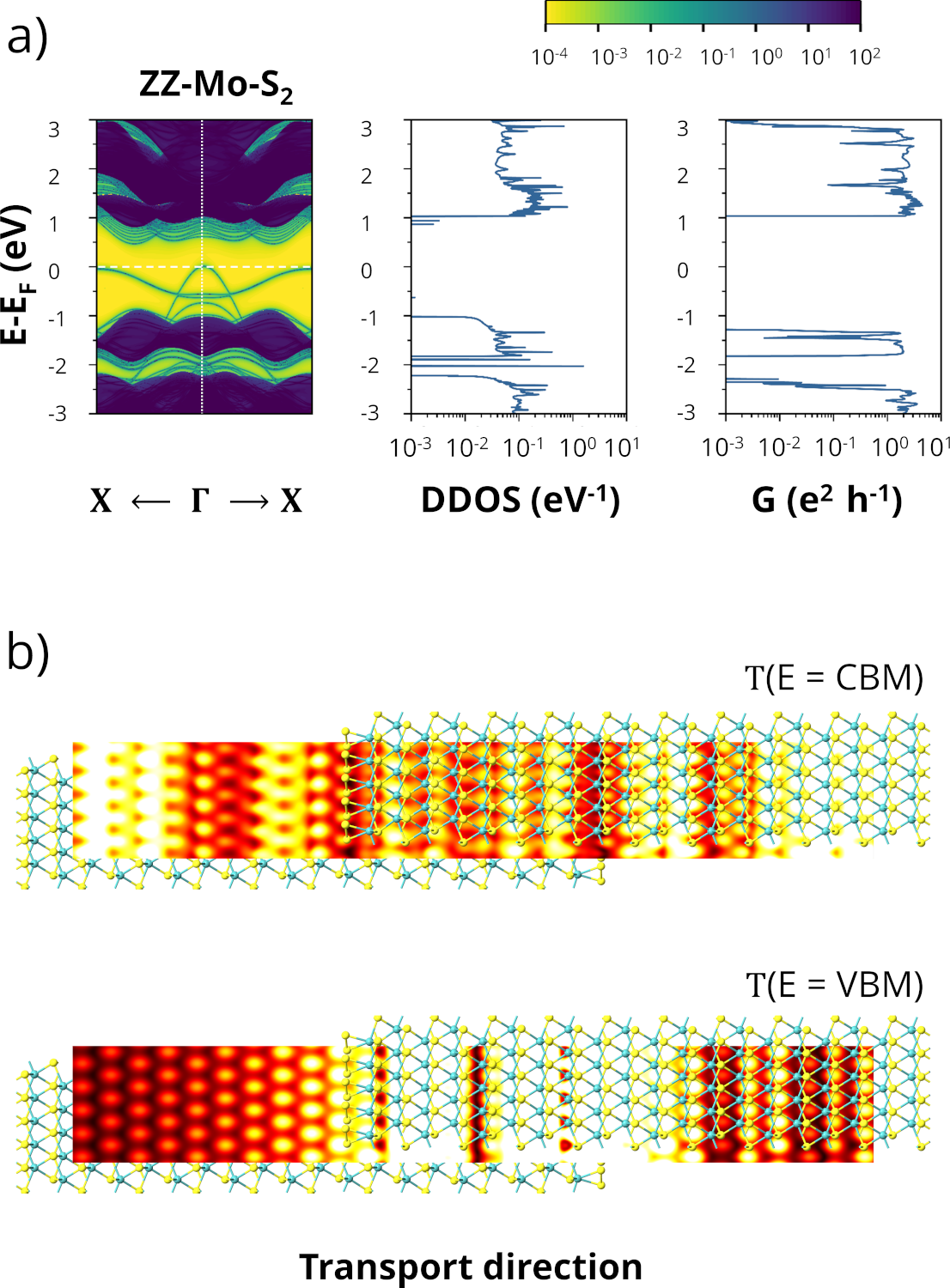}
    \caption{(a) Surface band structure, device density of states, and conductance as function of energy. (b) Transmission eigenstates at $E=CBM$ and $E=VBM$, with $k\textsubscript{a}=\Gamma$ for ZZ-Mo-S\textsubscript{2} device.}
    \label{fig_SI3}
\end{figure}

\begin{figure}[ht!]
    \centering
    \includegraphics[width=\textwidth]{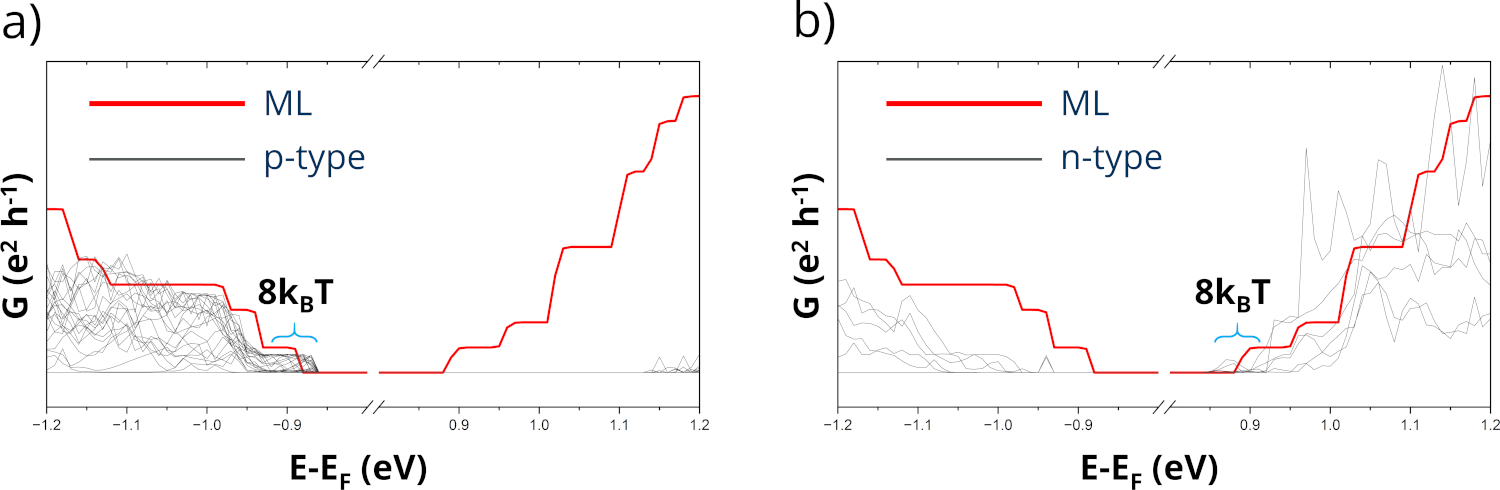}
    \caption{Conductance as function of energy of ML (red curve) compared to (a) p-type semiconductors (ZZ-S), and (b) n-type semiconductors (ZZ-Mo, ZZ-S-Mo, and ZZ-Mo-S\textsubscript{2}). The $\pm4k_{B}T$ energy window around CBM and VBM of ML denoted by blue curly bracket.}
    \label{fig_SI4}
\end{figure}

\begin{figure}[ht!]
    \centering
    \includegraphics[width=\textwidth]{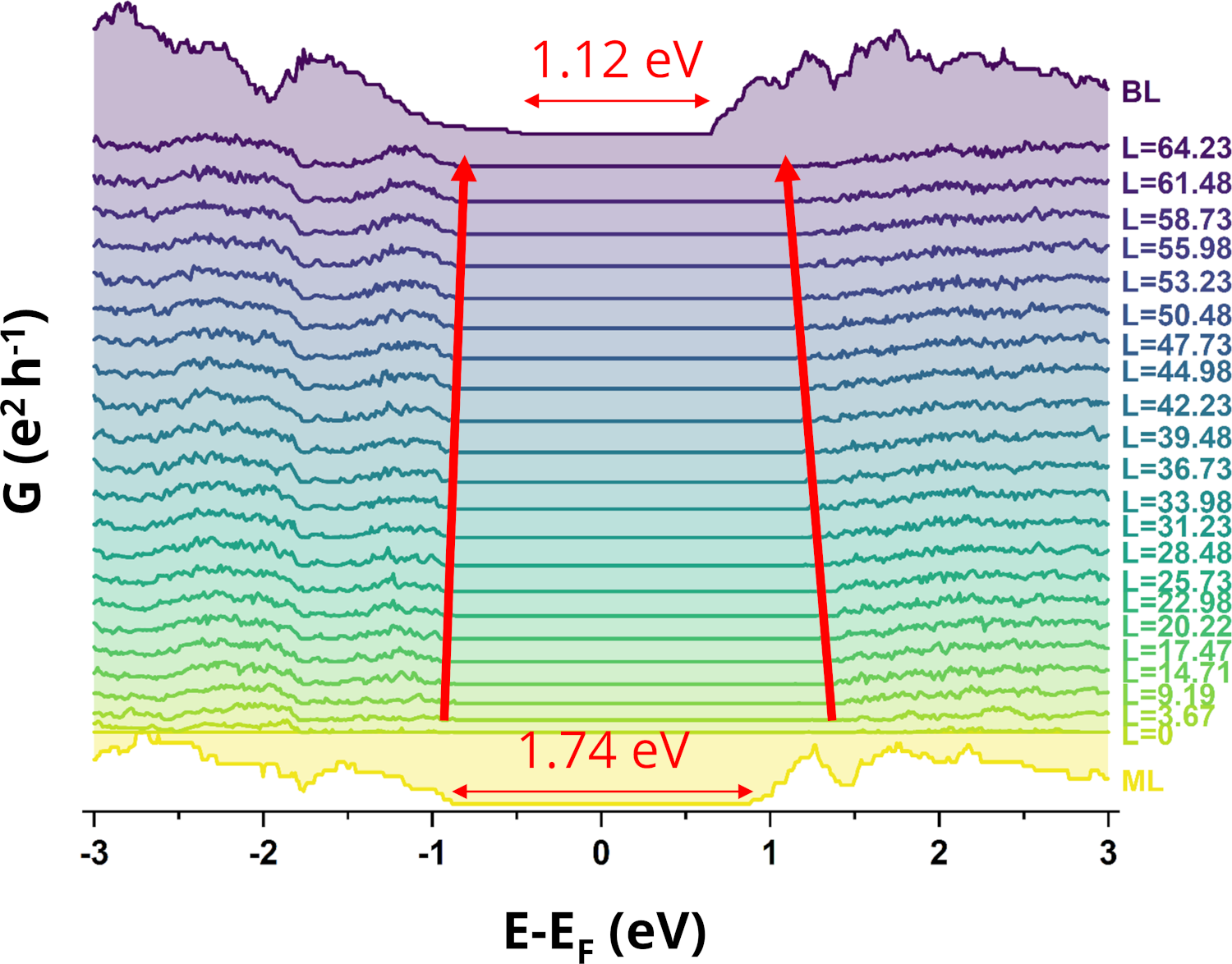}
    \caption{Conductance as function of energy of ZZ-S device with $L_{J}$ ranging from 0 to 64.23~\AA. The transport gap decreases with increasing $L_{J}$.}
    \label{fig_SI5}
\end{figure}

\begin{figure}[ht!]
    \centering
    \includegraphics[width=\textwidth]{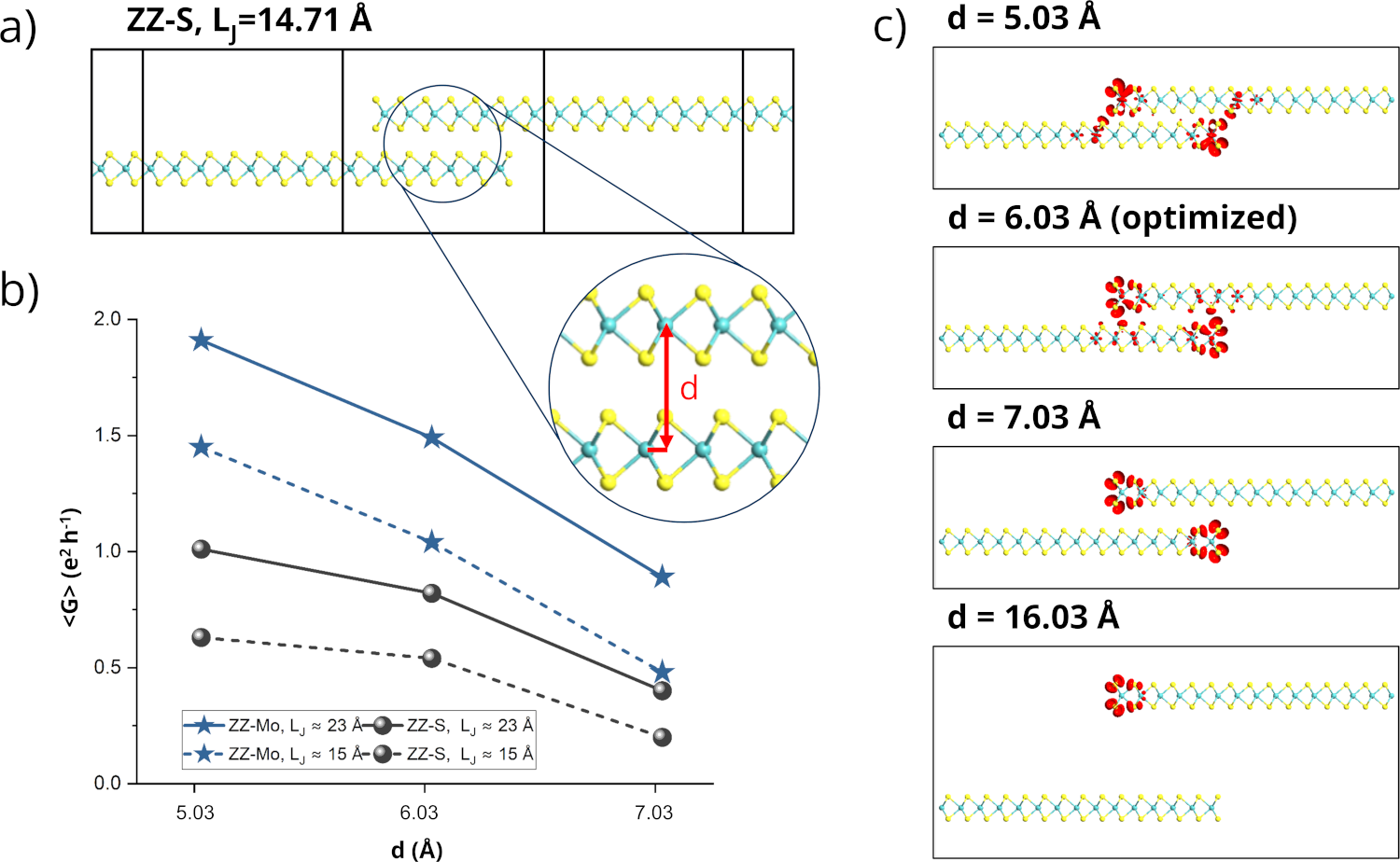}
    \caption{Analysis of the interlayer distance ($d$) impact on conductance and Bloch wave functions. (a) Vector representation of $d$. (b) Average conductance as function of $d$ for ZZ-S and ZZ-Mo. (c) Corresponding wave functions of Blcoh states for specific interlayer distances: $d=5.03$~\AA, 6.03~\AA (relaxed case), and 7.03~\AA, all with $L_{J}$ = 14.71~\AA\ for ZZ-S at the edges. The Bloch states are selected from edge states near the Fermi energy.}
    \label{fig_SI6}
\end{figure}

\begin{figure}[ht!]
    \centering
    \includegraphics[width=0.7\textwidth]{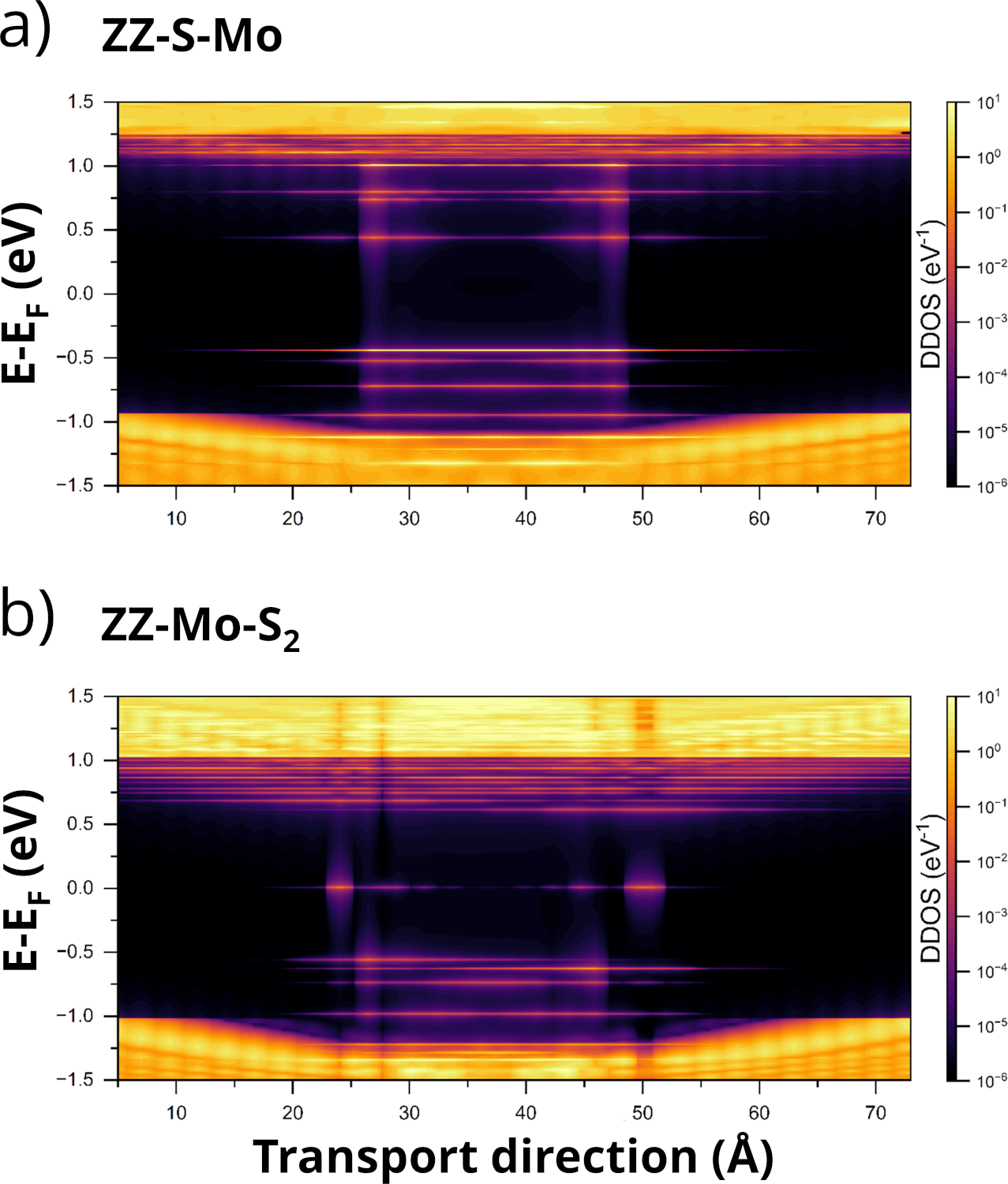}
    \caption{Local density of states as function of device length in (a) ZZ-S-Mo, and (b) ZZ-Mo-S\textsubscript{2}.}
    \label{fig_SI7}
\end{figure}

\begin{figure}[ht!]
    \centering
    \includegraphics[width=\textwidth]{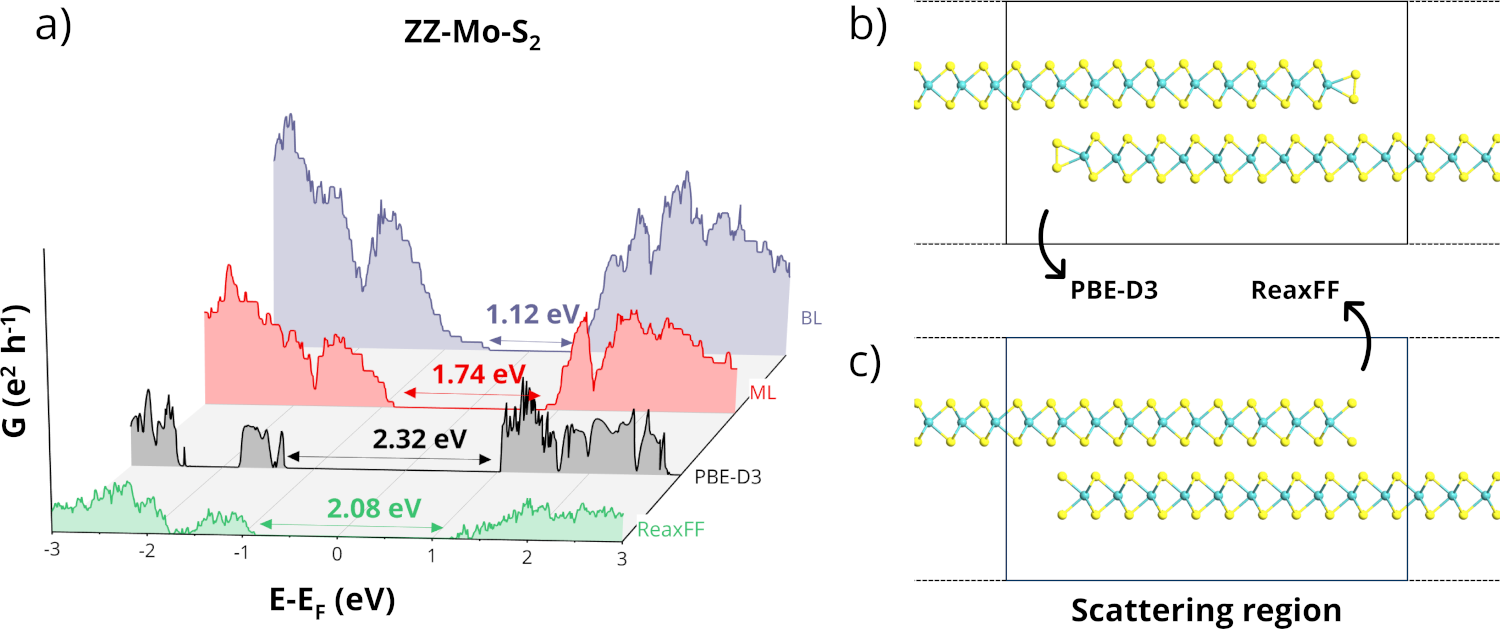}
    \caption{(a) Conductance as function of energy for ZZ-Mo-S\textsubscript{2}, with the scattering region fully optimized using PBE-D3 and ReaxFF methods. (b) Atomic configuration of the scattering region after optimization with PBE-D3, showing the formation of S\textsubscript{2} dimer at the edge. (c) Edges do not form dimer in ReaxFF optimization.}
    \label{fig_SI8}
\end{figure}

\end{document}